\documentclass[runningheads]{llncs}
\usepackage[utf8]{inputenc}
\usepackage{cite}
\usepackage{amsmath,amssymb,amsfonts}
\usepackage{algpseudocode}
\usepackage{graphicx}
\usepackage{pseudo}
\usepackage{textcomp}
\usepackage{xcolor}
\usepackage{enumitem}
\usepackage{theoremref}
\usepackage{todonotes}
\usepackage{comment}
\usepackage{mathrsfs}
\usepackage[colorlinks=true,linkcolor={blue},citecolor=blue]{hyperref}
\usepackage[noabbrev,capitalise,nameinlink]{cleveref}
\usepackage{microtype}
\usepackage{tikz}

\usetikzlibrary{arrows,shapes,automata,backgrounds,petri}
\usetikzlibrary{positioning,calc}
\usetikzlibrary{patterns}
\tikzstyle{legendborder}=[rectangle, draw, black, rounded corners, thin, top color=white, text=black, minimum width=2.5cm, text width=4.5cm]
\tikzstyle{legendnoborder}=[rectangle, draw, white, rounded corners, thin, top color=white, text=black, minimum width=2.5cm]
\tikzstyle{selected edge} = [draw,line width=1pt,black]
\usetikzlibrary{shapes,backgrounds,plothandlers,plotmarks,calc,arrows,fadings,decorations.pathreplacing,decorations.pathmorphing,}

\tikzstyle{v}=[circle,fill=black,draw=black!75,inner sep=0pt,minimum size=0.3em]
\tikzstyle{I}=[circle,draw=black!75,inner sep=0pt,minimum size=0.8em]
\tikzstyle{J}=[rectangle,draw=black!75,inner sep=0pt,minimum size=0.7em]
\tikzstyle{vertex}=[circle,inner sep=2,minimum size =2mm,semithick,fill=white!80!blue, draw=black]

\newcommand{\WW}{\textsf{W}}
\newcommand{\WT}{\textsf{W[2]}}
\newcommand{\WO}{\textsf{W[1]}}
\newcommand{\PP}{\textsf{P}}
\newcommand{\PPSPACE}{\textsf{PSPACE}}
\newcommand{\NPP}{\textsf{NP}}
\newcommand{\FPT}{\textsf{FPT}}

\newcommand{\Cc}{\mathscr{C}}
\newcommand{\Oof}{\mathcal{O}}

\title{A survey on the parameterized complexity of the  independent set and (connected) dominating set reconfiguration problems}
\titlerunning{Parameterized algorithms for reconfiguration problems}

\author{Nicolas Bousquet\inst{1}\thanks{Research supported by ANR project GrR (ANR-18-CE40-0032) and PHC Cedre project 2022 ``PLR''} \and Amer~E.~Mouawad\inst{2,3,4}\thanks{Research supported by
the Alexander von Humboldt Foundation, PHC Cedre project 2022 ``PLR'', and partially supported by URB project ``A theory of change through the lens of reconfiguration''.} \and Naomi Nishimura\inst{3}\thanks{Research supported by the Natural Sciences and Engineering Research Council of Canada.} \and\\ Sebastian Siebertz\inst{4}}
\authorrunning{N.~Bousquet, A.E.~Mouawad, N.~Nishimura, and S.~Siebertz}

\institute{CNRS, LIRIS, Universit\'e de Lyon, Universit\'e Claude Bernard Lyon 1, France \and American University of Beirut, Lebanon \and University of Waterloo, Canada \and University of Bremen, Germany}

\begin{document}
\maketitle

\begin{abstract}
A graph vertex-subset problem defines which subsets of
the vertices of an input graph are feasible solutions. We view a feasible solution as a set of tokens placed on the vertices of the graph. A reconfiguration variant of a vertex-subset problem asks, given
two feasible solutions of size~$k$,
whether it is possible to transform one into the other 
by a sequence of token slides (along edges of the graph) or token jumps (between arbitrary vertices of the graph) such that each intermediate set remains a feasible solution of size $k$. 
Many algorithmic questions present themselves in the form of reconfiguration problems: Given the description of an initial system state and the description of a target state, is it possible to transform the system from its initial state into the target one while preserving certain properties of the system in the process? Such questions have
received a substantial amount of attention under the so-called combinatorial reconfiguration framework.  
We consider reconfiguration variants of three fundamental underlying graph vertex-subset problems,
namely \textsc{Independent Set}, \textsc{Dominating Set}, and \textsc{Connected Dominating Set}. We survey both older and more recent work on the parameterized complexity of all three problems when parameterized by the number of tokens $k$. The emphasis will be on positive results and the most common techniques for the design of fixed-parameter tractable algorithms. 
\end{abstract}
\newpage
\section{Introduction}\label{sec-intro}
Given an $n$-vertex graph $G$ and two vertices $s$ and $t$ in $G$,
determining whether there exists a path between $s$ and $t$ is one of the most fundamental graph problems.
In the division between \PP\ versus \NPP\ or ``easy'' versus ``hard'', this problem, also known as the $(s,t)$-reachability problem, is on the easy side. That is, it can be solved in $poly(n)$ time,
where $poly$ is a polynomial function. 
But what if $G$ has~$2^n$ vertices?
To obtain a $poly(n)$-time algorithm, we can no longer assume $G$ to be part of the input, as reading the input
alone requires more than $poly(n)$ time. 
Instead, we are given an oracle
encoded using $poly(n)$ bits and
that can, in $poly(n)$ time, answer queries
of the form ``Is $u$ a vertex in $G$?'' or ``Is there an edge between $u$ and $v$?''.
Given such an oracle and two vertices of the $2^n$-vertex graph,
can we still determine if there is a path between $s$ and $t$ in $poly(n)$ time?

This seemingly contrived question naturally
appears in many practical and theoretical problems.
In particular, many algorithmic questions present themselves in the following form: Given the description of a system state and the description of a state we would prefer the system to be in, is it possible to transform the system from its current state into the more desired one without ``breaking'' certain properties of the system in the process?
This is exactly the type of questions 
asked under the combinatorial reconfiguration framework;  such questions, with generalizations and specializations, have
received a substantial amount of attention in recent years \cite{DBLP:journals/comgeo/BoseH09,DBLP:journals/siamcomp/GopalanKMP09,DBLP:journals/tcs/ItoDHPSUU11,DBLP:journals/comgeo/LubiwP15,DBLP:journals/dam/ItoKD12,DBLP:journals/algorithmica/MouawadN0SS17}. We refer the reader to the surveys by van den Heuvel~\cite{DBLP:books/cu/p/Heuvel13} and Nishimura \cite{DBLP:journals/algorithms/Nishimura18} for extensive background on combinatorial reconfiguration. 
Historically, the study of reconfiguration questions predates the field of computer science, as many classic one-player games can be formulated as reachability questions \cite{DBLP:journals/JS79,DBLP:journals/icga/KendallPS08}, e.g., the $15$-puzzle and Rubik's cube. Reconfiguration problems independently emerged in different areas such as graph theory \cite{DBLP:journals/jctb/VergnasM81,DBLP:journals/dm/CerecedaHJ08}, constraint satisfaction \cite{DBLP:journals/siamcomp/GopalanKMP09,DBLP:journals/siamdm/MouawadNPR17}, computational geometry \cite{DBLP:journals/comgeo/LubiwP15}, and even quantum complexity theory \cite{DBLP:journals/toct/GharibianS18}. 

Under the combinatorial reconfiguration framework, instead of trying to find a feasible solution to an instance
$\mathcal{I}$ of a graph vertex-subset problem $\mathcal{P}$, we investigate structural
and algorithmic questions related to the solution space of~$\mathcal{P}$.
Given an adjacency relation $\mathcal{A}$, defined
over feasible solutions of $\mathcal{I}$, and a solution size $k$, the solution space can be represented using a graph, 
called the {\em reconfiguration graph},
that contains one vertex for each feasible solution of size $k$ of $\mathcal{P}$ on instance $\mathcal{I}$ and an edge between any vertices whose associated
solutions are adjacent under $\mathcal{A}$. 
 Viewing a feasible solution as a set of tokens placed on the vertices of a graph $G$, we consider two natural (symmetric) adjacency relations.  For two feasible solutions $S$ and $S'$, we say a token \emph{slides} from $u \in V(G)$ to $v \in V(G)$ if $u \in S$, $v \not\in S$, $v \in S'$, $u \not\in S'$, and $\{u,v\} \in E(G)$. Similarly, a token \emph{jumps} from $u \in V(G)$ to $v \in V(G)$ if $u \in S$, $v \not\in S$, $v \in S'$, and $u \not\in S'$.  Using the adjacency relations of token sliding and jumping yields the {\em token sliding model} and {\em token jumping model}, respectively.
A walk in 
a reconfiguration graph generates the {\em reconfiguration sequence} formed out of the feasible solutions associated with the vertices in the walk. At times, we can, equivalently, view a reconfiguration sequence as the sequence of slides or jumps between consecutive pairs of solutions.

\begin{figure}[t]
\centering
    \begin{tikzpicture}[->, scale=.725, auto=left, remember picture,every node/.style={rectangle},inner/.style={rectangle},outer/.style={rectangle}]


    \node[outer] (DEGREE) at (0,0) [rectangle, fill=white, draw=black] {Bounded degree};
    \node[outer] (TREEWIDTH) at (11,1) [rectangle, fill=white, draw=black] {Bounded treewidth};
    \node[outer] (CLIQUEWIDTH) at (12,3) [rectangle, fill=white, draw=black] {Bounded cliquewidth};
    \node[outer] (TWINWIDTH) at (12,7) [rectangle, fill=white, draw=black] {Bounded twinwidth};
    \node[outer] (CHORDAL) at (11,0) [rectangle, fill=white, draw=black] {Chordal of bounded clique number};
    \node[outer] (PLANAR) at (5,0) [rectangle, fill=white, draw=black] {Planar};
    \node[outer] (GENUS) at (5,1) [rectangle, fill=white, draw=black] {Bounded genus};
    \node[outer] (MINOR) at (7,2) [rectangle, fill=white, draw=black] {$H$-minor free};
    \node[outer] (TOPMINOR) at (6,3) [rectangle, fill=white, draw=black] {$H$-topological-minor free};
    \node[outer] (EXPANSION) at (6,4) [rectangle, fill=white, draw=black] {Bounded expansion};
    \node[outer] (DENSE) at (9.5,5) [rectangle, fill=white, draw=black] {Nowhere dense};
    \node[outer] (DEGENERATE) at (2,5) [rectangle, fill=white, draw=black] {Bounded degeneracy};
    \node[outer] (BICLIQUE) at (6,6) [rectangle, fill=white, draw=black] {Biclique free};
    \node[outer] (SEMILADDER) at (6,7) [rectangle, fill=white, draw=black] {Semi-ladder free};

    \foreach \from/\to in {PLANAR/GENUS,
    CHORDAL/TREEWIDTH,
    TREEWIDTH/MINOR,
    GENUS/MINOR,
    MINOR/TOPMINOR,
    DEGREE/TOPMINOR, TREEWIDTH/CLIQUEWIDTH, CLIQUEWIDTH/TWINWIDTH, BICLIQUE/SEMILADDER}
    \draw[-stealth] (\from) -> (\to);
    \draw[->,-stealth,rounded corners] (MINOR.east) -- (9,3) -- (9,4) -- (12,4.5);

    \foreach \from/\to in {TOPMINOR/EXPANSION,
    EXPANSION/DENSE,
    DENSE/BICLIQUE,
    EXPANSION/DEGENERATE,
    DEGENERATE/BICLIQUE},
    \draw[-stealth] (\from) -> (\to);

    \end{tikzpicture}
\caption{Inclusion diagram of the mentioned graph classes. Arrows indicate inclusion.}
\label{fig-graph-classes}
\end{figure}
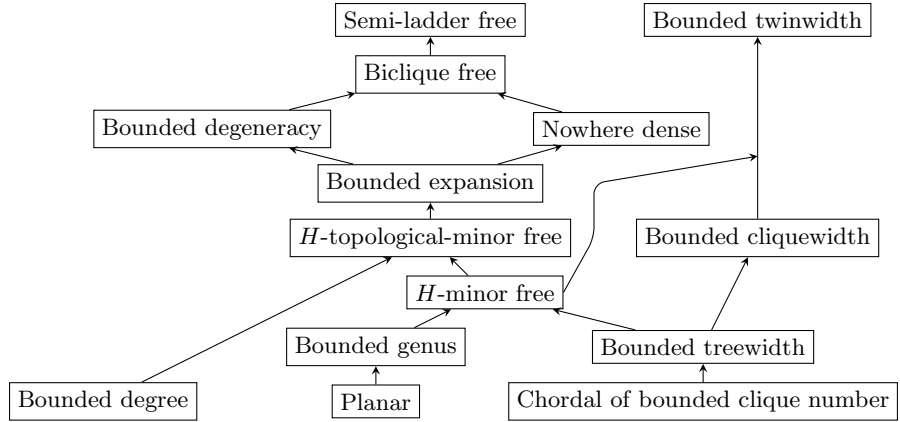

In this survey, we focus on the (parameterized complexity of) reconfiguration variants of three fundamental graph problems, namely {\sc Independent Set (IS)}, {\sc Dominating Set (DS)}, and {\sc Connected Dominating Set (CDS)}. Given a simple undirected graph $G$, a set of vertices $I \subseteq V(G)$ is an {\em independent set} if the vertices of $I$ are pairwise non-adjacent. Deciding whether a given graph contains an independent set of a given size $k$ is known to be \NPP-complete~\cite{DBLP:conf/coco/Karp72} and  \WO-complete parameterized by (solution size) $k$~\cite{DBLP:journals/siamcomp/DowneyF95}. A set of vertices $D \subseteq V(G)$ is a {\em dominating set} if every vertex in $V(G)$ is either in $D$ or has a neighbor in $D$. A dominating set $D$ is a {\em connected dominating set} if additionally the graph induced by $D$ is connected. Deciding whether a given graph contains a (connected) dominating set of a given size $k$ is known to be \NPP-complete~\cite{DBLP:conf/coco/Karp72} and \WT-complete parameterized by (solution size) $k$~\cite{DBLP:journals/siamcomp/DowneyF95}. On the positive side, a very fruitful line of research~\cite{DBLP:conf/icalp/Bodlaender88, DBLP:journals/jacm/GroheKS17, DBLP:journals/tcs/TelleV19, DBLP:journals/ejc/PilipczukS21} has shown that all three problems, i.e., \textsc{Independent Set}, \textsc{Dominating Set}, and {\sc Connected Dominating Set} parameterized by solution size $k$ become fixed-parameter tractable when restricted to sparse graph classes, such as  planar, bounded degree, bounded treewidth, nowhere dense, and biclique-free classes of graphs. For a monotone, i.e., subgraph-closed, class of graphs $\Cc$, it is easy to see that \textsc{Independent Set} is fixed-parameter tractable if and only if $\Cc$ excludes some
clique (a clique of size~$n$ contains every $n$-vertex graph as a subgraph) and \textsc{(Connected) Dominating Set} is fixed-parameter tractable
if and only if $\Cc$ excludes some biclique (a biclique with parts of size $n$ contains every bipartite graph with parts of
size $n$ as a subgraph). For dense graphs, 
the most general classes on which the problems are known to be
fixed-parameter tractable are classes of bounded cliquewidth~\cite{courcelle2000linear} (and more generally classes of bounded twinwidth when additionally a contraction sequence is given as part of the input~\cite{DBLP:conf/icalp/BonnetG0TW21,DBLP:journals/jacm/BonnetKTW22}) and classes of bounded semi-ladder index~\cite{DBLP:conf/stacs/FabianskiPST19} (generalizing biclique-free classes). It is therefore natural to ask whether the same is true, i.e., whether the same tractability results are true, for the reconfiguration variants of those underlying problems (when parameterized by the number of tokens) and this question will be the main focus of our survey. 

Before we go any further, let us first define the reconfiguration problems of interest. 

\begin{itemize}
    \item \textsc{Independent Set Reconfiguration} under token jumping ($\textsc{ISR-TJ}$)
    \item \textsc{Independent Set Reconfiguration} under token sliding ($\textsc{ISR-TS}$)
    \item \textsc{Dominating Set Reconfiguration} under token jumping ($\textsc{DSR-TJ}$)
    \item \textsc{Dominating Set Reconfiguration} under token sliding ($\textsc{DSR-TS}$)
    \item \textsc{Conn. Dom. Set Reconfiguration} under token jumping ($\textsc{CDSR-TJ}$)
    \item \textsc{Conn. Dom. Set Reconfiguration} under token sliding ($\textsc{CDSR-TS}$)
\end{itemize}

For each of the above problems, the input instance consists of a tuple $(G, k, S_s, S_t)$, where $G$ is a finite undirected simple graph, $k > 0$ is a positive integer, and $S_s$ and $S_t$ are the source and target feasible solutions of size~$k$. The goal is to decide whether $S_s$ can be transformed to $S_t$ by a sequence of jumps (under the token jumping model) or a sequence of slides (under the token sliding model). 

The {\sc ISR-TS} problem, introduced by Hearn and Demaine \cite{DBLP:journals/tcs/HearnD05}, has been extensively studied under the combinatorial reconfiguration framework~\cite{DBLP:conf/wg/BonamyB17,DBLP:conf/swat/BonsmaKW14,DBLP:conf/isaac/DemaineDFHIOOUY14,DBLP:conf/isaac/Fox-EpsteinHOU15,DBLP:conf/tamc/ItoKOSUY14,DBLP:journals/tcs/KaminskiMM12,DBLP:journals/jcss/LokshtanovMPRS18}. It is known that the problem is \PPSPACE-complete, even on restricted graph classes such as (planar) graphs of bounded bandwidth (and hence pathwidth)~\cite{DBLP:journals/jcss/Wrochna18,DBLP:conf/iwpec/Zanden15}, planar graphs of bounded degree~\cite{DBLP:journals/tcs/HearnD05}, split graphs~\cite{DBLP:journals/mst/BelmonteKLMOS21}, and bipartite graphs~\cite{DBLP:journals/talg/LokshtanovM19}. However, {\sc ISR-TS} can be decided in polynomial time on trees~\cite{DBLP:conf/isaac/DemaineDFHIOOUY14}, interval graphs~\cite{DBLP:conf/wg/BonamyB17}, bipartite permutation and bipartite distance-hereditary graphs~\cite{DBLP:conf/isaac/Fox-EpsteinHOU15}, and line graphs~\cite{DBLP:journals/tcs/ItoDHPSUU11}. 

The {\sc ISR-TJ} problem, introduced by Kami\'{n}ski et al.~\cite{DBLP:journals/tcs/KaminskiMM12}, is \PPSPACE-complete on (planar) graphs of bounded bandwidth~\cite{DBLP:journals/jcss/Wrochna18,DBLP:conf/iwpec/Zanden15} and planar graphs of bounded degree~\cite{DBLP:journals/tcs/HearnD05}. Lokshtanov
and Mouawad~\cite{DBLP:journals/talg/LokshtanovM19} showed that, unlike {\sc ISR-TS}, which is \PPSPACE-complete on bipartite graphs, the {\sc ISR-TJ} problem becomes \NPP-complete on bipartite graphs. 
On the positive side, it is ``easy'' to show that {\sc ISR-TJ} can be decided in polynomial-time on trees (and even on split/chordal graphs) since we can simply jump tokens 
to leaves (resp. vertices that only appear in the bag of a leaf in the clique tree) to transform one independent set into another.

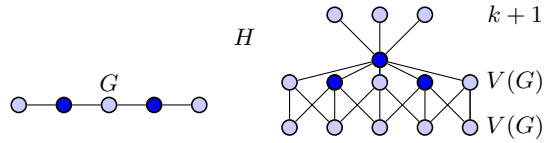
\begin{figure}
\begin{center}
\begin{tikzpicture}[scale=0.6]
\begin{scope}[yshift=0.5cm]
\node at (2,0.5) {$G$};
\node[vertex] (m0) at (0,0) {}; 
\node[vertex] (m1) at (1,0) {};
\node[vertex] (m2) at (2,0) {};
\node[vertex] (m3) at (3,0) {};
\node[vertex] (m4) at (4,0) {};
\draw[fill=blue] (1,0) circle (1.5mm);
\draw[fill=blue] (3,0) circle (1.5mm);
\draw[-] (m0) -- (m1) -- (m2) -- (m3) -- (m4);
\end{scope}

\begin{scope}[xshift=5cm]
\node[vertex] (n0) at (3,1.5){};

\foreach \i in {1,...,5}{
  \node[vertex] (a\i) at (\i,1){};
  \node[vertex] (b\i) at (\i,0){};
}
\foreach \i in {1,...,3}{
  \node[vertex] (n\i) at (\i + 1,2.5){};
}
\foreach \i in {1,...,5}{
  \draw (n0) -- (a\i);
  \draw (a\i) -- (b\i);
}
\foreach \i in {1,...,3}{
  \draw (n0) -- (n\i);
}
\draw[-] (a1) -- (b2) -- (a3) -- (b4) -- (a5) -- (b5) -- (a4) -- (b3) -- (a2) -- (b1);
\draw (n0) -- (a3);
\draw[fill=blue] (2,1) circle (1.5mm);
\draw[fill=blue] (4,1) circle (1.5mm);
\draw[fill=blue] (3,1.5) circle (1.5mm);

\node at (-0.2,0) {\textcolor{white}{$G$}};
\node at (6,0) {$V(G)$};
\node at (6,1) {$V(G)$};
\node at (6,2.5) {$k+1$};
\node at (0,2) {$H$};
\end{scope}
\end{tikzpicture}
\caption{A graph $G$ with a dominating set 
of size $k=2$ marked in dark blue and the graph $H$ 
obtained in the standard reduction from \textsc{Dominating 
Set} to \textsc{Connected 
Dominating Set}. $G$ has a dominating set of size $k$ if and
only if $H$ has a connected dominating set of size $k+1$. 
If $p$ is equal to the pathwidth of $G$, then the 
pathwidth of~$H$ is bounded by $2p+1$.}
\label{fig-ds-red}
\end{center}
\end{figure}

For the reconfiguration of dominating sets, earlier work focused on structural properties of the reconfiguration graph~\cite{DBLP:conf/isaac/BousquetJO20,DBLP:journals/gc/HaasS14,DBLP:journals/dm/HaasS17,DBLP:journals/jco/SuzukiMN16}. From an algorithmic standpoint, the authors  in~\cite{DBLP:journals/tcs/HaddadanIMNOST16} showed that \textsc{DSR-TJ} is \PPSPACE-complete on split graphs,  bipartite graphs, graphs of bounded 
bandwidth and planar graphs of maximum degree six. On the other hand, they gave linear-time algorithms for trees, interval graphs, and cographs. Bonamy et al.~\cite{DBLP:journals/dam/BonamyDO21} showed that \textsc{DSR-TS} is \PPSPACE-complete on split, bipartite and bounded treewidth graphs and polynomial-time solvable on dually chordal graphs and cographs. Bousquet and Joffard~\cite{DBLP:conf/fct/BousquetJ21} showed that the problem is polynomial-time solvable on circular-arc graphs and \PPSPACE-complete on circle graphs. Almost nothing is known for the \textsc{CDSR-TJ} and \textsc{CDSR-TS} problems except from what follows from the standard reduction from \textsc{DS} to \textsc{CDS}, which implies that \textsc{CDSR-TJ} is \PPSPACE-hard in most classes where \textsc{DSR-TJ} is \PPSPACE-hard (see~\cref{fig-ds-red}), e.g., graphs of bounded pathwidth.  

A systematic study of the parameterized complexity of
reconfiguration problems was initiated by Mouawad et al.~\cite{DBLP:journals/algorithmica/MouawadN0SS17}. This was followed by a long sequence of results trying to push the tractability limit of all aforementioned problems. It is the goal of this survey to collect and present these results in a comprehensive way. We cover the results for \textsc{ISR} in~\cref{sec-isr} and for \textsc{DSR} and \textsc{CDSR} in~\cref{sec-dsr}. Before doing so, we introduce all required background and terminology in~\cref{sec-prelim}. We conclude with open problems and directions for future research in \cref{sec-conclusion}.




\section{Preliminaries}\label{sec-prelim}
We denote the set of natural numbers by $\mathbb{N}$.
For $n \in \mathbb{N}$ we let $[n] = \{1, 2, \dots, n\}$.

\subsection{Graphs}
We assume that each graph $G$ is finite, simple, and undirected.
We let~$V(G)$ and $E(G)$ denote the vertex set and edge set of $G$, respectively. 
The {\em open neighborhood} of a vertex $v$ is denoted by $N_G(v) = \{u \mid \{u,v\} \in E(G)\}$ and the {\em closed neighborhood} by $N_G[v] = N_G(v) \cup \{v\}$. 
For a set of vertices $Q \subseteq V(G)$, we define $N_G(Q) = \bigcup_{v \in Q} N_G(v) \setminus Q$ and $N_G[Q] = N_G(Q) \cup Q$. 
The subgraph of~$G$ \emph{induced} by $Q$ is denoted by $G[Q]$, where $G[Q]$ has vertex set~$Q$ and edge set $\{\{u,v\} \in E(G) \mid u,v \in Q\}$. 
We let $G - Q = G[V(G) \setminus Q]$.

A {\em walk} of length $\ell$ from $v_0$ to $v_\ell$ in $G$ is a vertex sequence $v_0, \ldots, v_\ell$, such that for all $i \in \{0, \ldots, \ell-1\}$, $\{v_i,v_{i + 1}\} \in E(G)$.
It is a {\em path} if all vertices are distinct. 
It is a {\em cycle} if $\ell \geq 3$, $v_0 = v_\ell$, and $v_0, \ldots, v_{\ell - 1}$ is a path.
For a pair of vertices~$u$ and~$v$ in $V(G)$, by $\textsf{dist}_G(u,v)$ we denote the {\em distance} between $u$ and $v$ in $G$ (which is the length of a shortest path between $u$ and $v$ measured in number of edges and set to $\infty$ if $u$ and $v$ belong to different connected components).
The \emph{girth} of $G$, $\textsf{girth}(G)$, is the length of a shortest cycle contained in $G$. The girth of an acyclic graph (i.e.\ a forest) is defined to be infinity.

A class $\Cc$ of graphs is \emph{monotone} if it is closed
under taking subgraphs, i.e., if $G\in \Cc$ and $H\subseteq G$,
then also $H\in \Cc$. It is \emph{hereditary} if it is closed
under taking induced subgraphs. We do not distinguish between
isomorphic graphs. 

For $A,B\subseteq V(G)$ we write $E(A,B)$ for the set of edges with one endpoint in~$A$ and one endpoint in $B$.
For disjoint subsets $A,B$ of $V(G)$, we write $G[A,B]$ for the subgraph of $G$ \emph{semi-induced} by $A$ and $B$, that is, the subgraph with vertex set $A \cup B$ and all the edges with one endpoint in $A$ and one endpoint in $B$. 
A bipartite graph $H$ is a semi-induced subgraph of $G$ if $H = G[A,B]$ for some disjoint subsets $A$ and $B$ of $V(G)$. 
Let $H$ be a bipartite graph with vertices 
$a_1,\ldots, a_n$ (in one part) and $b_1,\ldots, b_n$ (in the other part). Then $H$ is:
\begin{itemize}
\item a \emph{biclique} of order $n$ if we have $\{a_i, b_j\}\in E(H)$ for all $i,j\in [n]$,
\item a \emph{co-matching} of order $n$ if we have $\{a_i,b_j\}\in E(H)\Longleftrightarrow i\neq j$ for all $i,j\in [n]$, 
\item a \emph{ladder} of order $n$ if $\{a_i,b_j\}\in E(H) \Longleftrightarrow i\leq j$ for all $i,j\in [n]$, and
\item a \emph{semi-ladder} of order $n$ if $\{a_i, b_j\} \in E(H)$ for all $i, j \in [n]$ with $i > j$, and $\{a_i, b_i\} \not\in E(H)$ for all $i \in [n]$.
\end{itemize}
The \emph{biclique index}, \emph{co-matching index}, \emph{ladder index}, or  \emph{semi-ladder index} of a graph $G$ is 
the largest $n$ such that
$G$ contains a biclique, co-matching, ladder, or semi-ladder of order $n$ as a semi-induced subgraph, respectively. 
A class of graphs is
\emph{biclique free},
\emph{co-matching free}, \emph{ladder free}, or \emph{semi-ladder free} if 
the supremum of the biclique indices, co-matching indices, ladder indices, or semi-ladder indices of its members is finite, respectively. 

Let $H = (V, E)$ be a hypergraph. A set $X$ of vertices of $H$ is \emph{shattered} if for every subset $Y$ of $X$ there exists a hyperedge $e$ such that $e \cap X = Y$. An intersection
between $X$ and a hyperedge $e$ of $E$ is called a \emph{trace (on $X$)}. Equivalently, a set $X$ is shattered if all its $2^{|X|}$ traces exist (in $H$). The \emph{Vapnik-Chervonenkis dimension} (\emph{VC-dimension}, for short) of a hypergraph is the maximum size of a shattered set. 

Let $G = (V, E)$ be a graph. The \emph{closed neighborhood hypergraph of $G$} is the hypergraph with vertex set $V(G)$ and where $X \subseteq V(G)$ is a hyperedge if and only if
$X = N[v]$ for some vertex $v \in V$. The VC-dimension of a graph is the VC-dimension of its closed neighborhood
hypergraph. The VC-dimension of a class of graphs $\Cc$ is the maximum VC-dimension of a graph of $\Cc$.

For a graph $G$ and a set $X \subseteq V(G)$, we often partition the vertices of $V(G) \setminus X$ into what we call \emph{$X$-projection classes} (we write \emph{projection classes} when $X$ is clear from context). That is, all vertices $u,v \in V(G) \setminus X$ of one class satisfy $N(u) \cap X = N(v) \cap X$. For $Y \subseteq X$, we let $\mathcal{C}_Y$ denote the set of vertices of $V(G) \setminus X$ whose neighborhood in $X$ is exactly $Y$. 

\subsection{Parameterized complexity}
A problem is {\em fixed-parameter tractable}, \FPT\ for short, on a class $\Cc$ of graphs with respect to a parameter $k$, if there is an algorithm deciding whether a given graph $G \in \Cc$ admits a solution of size $k$ in time $f(k) \cdot |V(G)|^c$, for a computable function~$f$ and constant $c$.  A {\em kernelization algorithm} is a polynomial-time algorithm that reduces an input instance to an equivalent instance of size bounded in the parameter only (independent of the input graph size), known as a {\em kernel}; we will say that two instances are {\em equivalent} if they are both yes-instances or both no-instances.  Every fixed-parameter tractable problem admits a kernel, however, possibly of exponential or worse size. For efficient algorithms it is therefore most desirable to obtain polynomial, or even linear, kernels.  The {\em \WW-hierarchy} is a collection of parameterized complexity classes $\FPT \subseteq \WW[1] \subseteq \ldots \subseteq \WW[t]$. The conjecture $\FPT \subsetneq \WW[1]$ can be seen as the analogue of the conjecture that $\PP \subsetneq \NPP$.  Therefore, showing hardness in the parameterized setting is usually accomplished by establishing an \FPT-reduction from a \WO-hard problem. We refer to the textbooks~\cite{DBLP:books/sp/CyganFKLMPPS15,DBLP:series/mcs/DowneyF99} for extensive background on parameterized complexity.

Hence, for a reconfiguration problem, a kernelization algorithm is a polynomial-time algorithm that transforms every input instance $(G,k,S_s,S_t)$ into an instance $(G',k',S'_s,S'_t)$ with $|G'| + k' \leq f(k)$ for some function $f$ and such that $(G,k,S_s,S_t)$ is a yes-instance if and only if $(G',k',S'_s,S'_t)$ is a yes-instance. On the reduced instance $(G',k',S'_s,S'_t)$ one can then run a brute-force algorithm to decide whether the initial instance was a yes-instance.

\subsection{Combinatorial reconfiguration}

As noted in the introduction, each reconfiguration sequence can be viewed either as a sequence of feasible solutions (and associated vertices in the reconfiguration graph) or as a sequence of token slides or jumps. Natural choices of parameters include the number $k$ of tokens (or, equivalently, size of feasible solutions $k$) as well as the length $\ell$ of a reconfiguration sequence. Although not the main focus of this survey, research on reconfiguration may consider not only the question of whether or not there exists a reconfiguration sequence between two feasible solutions, but also the length of such a sequence, as well as other properties of the reconfiguration graph.




The concept of irrelevant vertices from parameterized
complexity is adapted to introduce the notions of irrelevant and strongly irrelevant vertices for reconfiguration. 
For any vertex-subset problem $\mathcal{P}$, $n$-vertex graph $G$, positive integer~$k$,
and feasible solutions $S$ and $S'$,
we say a vertex $v \in V(G)$ is {\em irrelevant} 
(with respect to $S$ and $S'$)
if and only if whenever there exists a reconfiguration sequence from $S$ to $S'$ 
in the reconfiguration graph for $G$ and solution size $k$, then there exists a reconfiguration sequence from $S$ to $S'$ 
in the reconfiguration graph for $G - v$ and solution size $k$.
We say $v$ is {\em strongly irrelevant} (with respect to $S$ and $S'$) if 
the deletion of $v$ in $G$ does not modify the distance between the vertices corresponding to $S$ to $S'$ in the reconfiguration graph.
At a high level, it is enough to ignore irrelevant vertices when trying to find {\em any}
reconfiguration sequence between two feasible solutions, but only
strongly irrelevant vertices can be ignored if we wish to find a {\em shortest}
reconfiguration sequence. 

\section{Reconfiguration of independent sets}\label{sec-isr}

\subsection{Independent set reconfiguration under token jumping}

On general graphs, $\textsc{ISR-TJ}$ is $\WO$-hard when parameterized by the number of tokens $k$~\cite{DBLP:journals/algorithmica/MouawadN0SS17} (even when parameterized by $k + \ell$). The reduction is from the \textsc{Independent Set} problem parameterized by solution size $k$. Given an instance $(G,k)$ of \textsc{Independent Set}, we construct an instance $(G + K_{k+1,k+1}, k + 1, I_s, I_t)$ of $\textsc{ISR-TJ}$, where $G + K_{k+1,k+1}$ is the graph consisting of the disjoint union of~$G$ with a biclique having $k + 1$ vertices in each part. We let $L$ and $R$ denote the two parts of the biclique and we set $I_s = L$ and $I_t = R$. It is not hard to see that $(G,k)$ is a yes-instance of \textsc{Independent Set} if and only if $(G + K_{k+1,k+1}, k + 1, I_s, I_t)$ is a yes-instance of $\textsc{ISR-TJ}$; before any token can jump to $R$ we must have at least $k$ tokens in $G$ forming an independent set. $\textsc{ISR-TJ}$ remains $\WO$-hard even restricted to $\{C_4, \ldots, C_p\}$-free graphs~\cite{DBLP:journals/algorithmica/BartierBDLM21}, for any $p \geq 4$.
Having established hardness, we now focus on the parameterized complexity of $\textsc{ISR-TJ}$ on sparse classes of graphs.


\subsubsection{Graphs of bounded degree.} We first explain why $\textsc{ISR-TJ}$ is fixed-parameter tractable for the easy case of graphs $G$ of maximum degree $\Delta$~\cite{DBLP:journals/algorithmica/BartierBDLM21}. Let $I_s,I_t$ be the source and target independent sets. Since $I_s \cup I_t$ has at most $2k \Delta$ neighbors, there are two options, depending on the number of vertices in the graph. If $G$ has at most $3k(\Delta+1)$ vertices, then we can construct the reconfiguration graph via brute force in \FPT-time. On the other hand, if $G$ has more than $3k(\Delta+1)$ vertices,  then $V(G) \setminus N[I_s \cup I_t]$ contains an independent set $J$ of size at least~$k$. It is then possible to form a reconfiguration sequence by moving the tokens in $I_s$ into $J$ one token at a time and then moving tokens from $J$ into $I_t$ in the same manner. Thus, $\textsc{ISR-TJ}$ parameterized by $k$ admits a kernel of linear size on classes of graphs of bounded degree.

In more general sparse classes of graphs, a reconfiguration sequence might not exist even for arbitrarily large graphs since there might exist (almost) universal vertices, i.e., vertices connected to (almost) all. However, due to the sparsity constraints, very few such vertices can exist. As we shall see, in many cases, e.g., planar, nowhere dense, as well as $K_{h,h}$-free graphs, one can use such vertices to find one or more irrelevant vertices. 



\subsubsection{Planar graphs and $K_{h,h}$-free graphs.} Another typical sparse graph class is the class of planar graphs. Ito et al.~\cite{DBLP:conf/isaac/ItoKO14} proved the following:

\begin{theorem}\label{thm:ISR-TJ_planar}
$\textsc{ISR-TJ}$ parameterized by $k$ is fixed-parameter tractable on the class of planar graphs.
\end{theorem} 

We now sketch the main components of the proof of~\cref{thm:ISR-TJ_planar}. For $I_s$ and~$I_t$ the source and target independent sets, respectively, we consider projection classes of neighborhoods in $X= I_s \cup I_t$, i.e., $X$-projection classes, 
and prove either the existence of a yes-instance or that each $X$-projection class can be reduced to having at most $f(k)$ vertices, for some computable function $f$. Since $|I_s \cup I_t| \leq 2k$ there are at most $2^{2k} = 4^k$ projection classes\footnote{Actually one can prove that planar graphs have a linear number of classes.}, and we therefore obtain the desired kernel.

Recall that for $Y \subseteq X$, $\mathcal{C}_Y$ denotes the set of vertices $y$ of $V(G) \setminus X$ such that $N(y) \cap X = Y$. The vertices of $\mathcal{C}_Y$ form the \emph{$Y$-class}; a $Y$-class is a \emph{$(Y,r)$-class} when $|Y|=r$. Since planar graphs are $K_{3,3}$-free, no three vertices can share more than two neighbors, and hence the number of vertices in a $(Y,r)$-class with $r \ge 3$ is at most $2$.

As every planar graph is $4$-colorable, every subgraph of size at least $4k$ contains an independent set of size at least $k$. Thus, if any $(Y,r)$-class where $r \le 1$ is large enough, we can simply transform $I_s$ into an independent set $J$ contained in $\mathcal{C}_Y$ and then transform $J$ into $I_t$; adding as possible first and last steps the jumps of the token in $I_s \cap Y$ and the token in $I_t \cap Y$, if those tokens exist.

To complete the proof, it suffices to consider $(Y,2)$-classes.  Ito, Kamiński, and Ono~\cite{DBLP:conf/isaac/ItoKO14} proved that if $\mathcal{C}_Y$ is large enough, then it can be replaced by an independent set of size $k$. Suppose that at some point of the reconfiguration sequence a token is moved to a vertex in $\mathcal{C}_Y$ to form an independent set $I$. Clearly, $I$ cannot contain any vertex in $Y$. In addition, the fact that $G$ does not contain $K_{3,3}$ ensures that every vertex of $V(G) \setminus Y$ has at most two neighbors in $\mathcal{C}_Y$ (recall that $I$ does not intersect with $Y$ and $|\mathcal{C}_Y|\geq 3$). In particular, no vertex in $I$ can have more than two neighbors in $\mathcal{C}_Y$, so that the set $I$ has at most $2k$ neighbors in~$\mathcal{C}_Y$. Thus, for large enough $\mathcal{C}_Y$, we can always find a vertex in $\mathcal{C}_Y$ that is not in the neighborhood of the current independent set $I$. Hence, when~$\mathcal{C}_Y$ is large enough we can instead retain an independent set of size $k$.


Given that we can bound the size of all the classes one can easily prove that we get a kernel of polynomial size. One can then ask the following question:

\begin{question}
Does $\textsc{ISR-TJ}$ admit a linear kernel on planar graphs?
\end{question}



The intuition behind the proof of \cref{thm:ISR-TJ_planar}
for $K_{3,3}$-free graphs is that $(Y,r)$-classes for $r \geq 3$ are of bounded size and $(Y,r)$-classes for $r \leq 2$ are either of bounded size, or can be reduced to bounded size in forming a kernel. However, when we consider $K_{4,4}$-free graphs (or more generally $K_{h,h}$-free graphs), there can exist a $(Y,2)$-class $\mathcal{C}_Y$ that does not immediately imply a yes-instance (in the sense that we cannot guarantee that we can immediately move source and target independent sets to an independent set of $\mathcal{C}_Y$) nor is easily reducible (in the sense that a vertex of $V(G) \setminus Y$ can be adjacent to arbitrarily many vertices of $\mathcal{C}_Y$). 

Note that if no vertex is adjacent to many vertices in $\mathcal{C}_Y$, then we can replace the $Y$-class by an independent set of size $k$, just as in the case of $K_{3,3}$-free graphs.
Otherwise, a result of K\"{o}v\'ari, S\'os, Tur\'an \cite{KovariST54} ensures that in $K_{h,h}$-free graphs, for every $\epsilon >0$, the number of vertices $Z_Y$ incident to an $\epsilon$-fraction of the vertices of the set $\mathcal{C}_Y$ is bounded (in terms of $\epsilon$ and $h$).



The key idea of~\cite{DBLP:conf/fct/BousquetMP17} to adapt the proof for $K_{h,h}$-free graphs entails updating the set $X=I_s \cup I_t$ by adding to $X$  the vertices of $Z_Y$ for every $Y \subseteq X$, thereby forming the set $X'$. We can now refine the categorization of the projection classes depending on their neighborhoods in $X'$. Let $Y' \subseteq X'$ such that the $Y'$-class is large.  Either $Y' \nsubseteq X$ and then the $Y'$-class is a refinement of a $Y$-class where $Y \subseteq X$ and $|Y| < |Y'|$. Since $(Y',h)$-classes have size at most $h-1$, we intuitively ``gained'' something since we cannot increase too often the neighborhood of a class in $X$. But if $Y' \subseteq X$ we did not gain anything. 

The algorithm proposed in~\cite{DBLP:conf/fct/BousquetMP17} consists of repeating this refinement\footnote{The exact refinement is actually slightly more involved but the following paragraph, though not entirely accurate, supplies the intuition behind the proof.} by  iteratively defining sets $X_1=X,X_2,X_3,X_4,\ldots,X_r$ (where $r$ only depends on $k$ and $h$). The authors finally prove that if, after all these steps, the size of a class is large enough, then replacing this class by an independent set of size $k$ does not modify the existence of a reconfiguration sequence from $I_s$ to $I_t$. In other words, even if the neighborhood in $X_r$ of a class has not increased compared to~$X$, then we can nevertheless reduce it. 



\subsubsection{Degenerate and nowhere dense graphs.} Even if degenerate and nowhere dense graphs are included in biclique-free graphs, we think that it is worth mentioning the proof techniques of~\cite{DBLP:journals/jcss/LokshtanovMPRS18}, which are of independent interest.

For both graph classes, the authors in~\cite{DBLP:journals/jcss/LokshtanovMPRS18} show that one can find irrelevant vertices by making use of the classical result of Erd\H{o}s and Rado~\cite{ER60}, also
known in the literature as the sunflower lemma. A \emph{sunflower} with $r$ \emph{petals} and a \emph{core}~$Y$ is a collection of sets $S_1, \ldots, S_r$ such that $S_i \cap S_j = Y$ for all $i \neq j$; the sets $S_i \setminus Y$ are petals and we require none of them to be empty. Note that a family of pairwise
disjoint sets is a sunflower (with an empty core). Now assume that one can find a large (in terms of the number of tokens $k$) sunflower with core $Y$ in the collection of sets defined by the closed neighborhoods of a set of vertices $v_1, \ldots, v_q$ in the graph. That is, we have $N[v_1], \ldots, N[v_p]$, $p \leq q$, such that such that $N[v_i] \cap N[v_j] = Y$ for all $i \neq j$. Note that if any independent set $I$ of size $k$ intersects with the core then $I$ cannot contain any vertex from $v_1, \ldots, v_p$. Otherwise, $I$ can intersect with at most $k$ petals. Moreover, any vertex of the graph which is not included in the core or in some petal cannot be adjacent to a vertex from $v_1, \ldots, v_p$. Hence, if $p$ is large enough (in terms of $k$), then we can always delete one of the vertices from $v_1, \ldots, v_p$ without affecting the existence of a reconfiguration sequence from $I_s$ to $I_t$. This follows from the fact that we can always find another vertex in $v_1, \ldots, v_p$ that will be used to ``mimic'' the behavior of the deleted vertex. 

To find large sunflowers in degenerate graphs one can immediately apply the sunflower lemma; in a degenerate graph
at least half of the vertices have degree at most twice the degeneracy (since the average degree of a $d$-degenerate graph is at most $2d)$. Hence, as long as the graph is large enough, one can always find a sunflower of the appropriate size and delete one irrelevant vertex. 



The algorithm for nowhere dense graphs closely mimics the previous algorithm in the following sense. Instead of using the sunflower lemma to find a large sunflower, the authors use 
the so-called notion of uniform quasi-wideness~\cite{DBLP:conf/fsttcs/DawarK09} to find a ``large enough almost sunflower'' with an initially ``unknown'' core and then use structural properties of the graph to find this core and complete the sunflower. 
At a high level, uniform quasi-wideness states that if $G$ comes from a nowhere dense class and $A \subseteq V(G)$ is large enough, then we can find a small set $X \subseteq V(G)$ whose deletion leaves a large set $B \subseteq A$ that is $2$-independent in $G - X$ (where a set $B$ is $2$-independent whenever its vertices are pairwise non-adjacent and pairwise do not share any common neighbors). The trick consists of first looking at $(I_s \cup I_t)$-projection classes and then finding a large class in which we can find the sets $B$ and $X$ (using uniform quasi-wideness). Then, we further classify the vertices of $B$ into $X$-projection sub-classes. The petals of the sunflower, which consist of vertices of $B$ and their neighbors (recall that $B$ is $2$-independent) can then be found in a large $X$-projection sub-class and the core will be a subset of $X \cup I_s \cup I_t$. The existence of a large $X$-projection sub-class can be guaranteed by appropriately choosing the sizes of $B$ and $X$.

\subsubsection{The curious case of bipartite graphs.} Although bipartite graphs are not sparse nor appear in~\cref{fig-graph-classes}, they merit attention, since we believe that there remain several interesting questions that have yet to be answered. It is still unknown if $\textsc{ISR-TJ}$ is \WO-hard on bipartite graphs. However, Agrawal et al.~\cite{DBLP:conf/iwpec/AgrawalAD21} showed that the problem is unlikely to be fixed-parameter tractable.
The proof is based on the fact that \textsc{Balanced Biclique} does not admit an \FPT-time $2$-approximation algorithm assuming Gap-ETH\footnote{Informally speaking, Gap-ETH states that we cannot, for some $\epsilon >0$, distinguish in subexponential time $3$-SAT formulas that are satisfiable from those which are $\epsilon$-far from being satisfiable.}~\cite{DBLP:journals/siamcomp/ChalermsookCKLM20}.
Equivalently, we cannot distinguish in \FPT-time (parameterized by $k$) a graph that admits a balanced biclique of size $k$ from a graph that does not admit a balanced biclique of size~$k/2$. 

The proof of~\cite{DBLP:conf/iwpec/AgrawalAD21} simply consists of constructing a ``weak'' reduction from \textsc{Balanced Biclique} to $\textsc{ISR-TJ}$ on bipartite graphs with the following properties:
\begin{itemize}
    \item if we have a yes-instance of \textsc{Balanced Biclique}, then we have a yes-instance of $\textsc{ISR-TJ}$ and,
    \item if we have a yes-instance of $\textsc{ISR-TJ}$, then the original graph admits a balanced biclique of size $k/2$.
\end{itemize}
A fixed-parameter tractable algorithm for $\textsc{ISR-TJ}$ on bipartite graphs would imply that we could distinguish in \FPT-time graphs having balanced bicliques of size $k$ from graphs having balanced bicliques of size less than $k/2$, a contradiction under Gap-ETH.


\subsubsection{More open problems.}
Probably the most exciting research direction is related to dense graph classes on which almost nothing is known. In particular one might ask the following:

\begin{question}
Is $\textsc{ISR-TJ}$ parameterized by $k$ fixed-parameter tractable on graphs of bounded semi-ladder index?
\end{question}

\begin{question}
Is $\textsc{ISR-TJ}$ parameterized by $k$ fixed-parameter tractable on graphs of bounded cliquewidth (and, more generally, bounded twinwidth)?
\end{question}

We conclude this section with one more interesting open question. There is a correlation between VC-dimension and complete bipartite subgraphs. Namely, a $K_{h,h}$-free graph has VC-dimension at most $\Oof(h)$. Since the $\textsc{ISR-TJ}$ problem is \WO-hard on general graphs and fixed-parameter tractable on
$K_{h,h}$-free graphs, one can naturally ask if this result can be extended to graphs of bounded VC-dimension. It was shown in~\cite{DBLP:conf/fct/BousquetMP17} that $\textsc{ISR-TJ}$ is polynomial-time solvable on graphs of VC-dimension $1$, \NPP-hard on graphs of VC-dimension $2$, and \WO-hard on graphs of VC-dimension $3$. The following question remains open:

\begin{question}
Is $\textsc{ISR-TJ}$ parameterized by $k$ fixed-parameter tractable on graphs of VC-dimension $2$?
\end{question}



\subsection{Independent set reconfiguration under token sliding}
Just like the token jumping variant, $\textsc{ISR-TS}$ is \WO-hard on general graphs. It remains hard even when restricted to bipartite graphs or $\{C_4, \ldots, C_p\}$-free graphs~\cite{DBLP:journals/algorithmica/BartierBDLM21}, for any $p \geq 4$. We sketch the proof for general graphs, which mimics the reduction for the token jumping model. The reduction is from the \textsc{Multicolored Independent Set (MIS)} problem, known to be \WO-hard. In the \textsc{Multicolored Independent Set} problem, we are given a graph consisting of $k$ cliques of arbitrary size and extra edges joining vertices from different cliques. The goal is to find an independent set of size $k$ which must intersect with each clique in exactly one vertex (called a multicolored independent set). 
Given an instance $(G = (V_1 \cup 
\ldots \cup V_k, E),k)$ of \textsc{Multicolored Independent Set}, we construct an instance $(H, k, I_s, I_t)$ of $\textsc{ISR-TS}$ as follows. We first let $H = G + K_{k,k}$ where without loss of generality each set $V_i$ can be assumed to be a clique. We let $L = \{l_1, \ldots, l_k\}$ and $R = \{r_1, \ldots, r_l\}$ denote the two parts of the biclique, and we set $I_s = L$ and $I_t = R$. Finally, we add edges between $\{l_i,r_i\}$ and every vertex in $V_i$, for $i \in [k]$. It is not hard to see that $(G,k)$ is a yes-instance of \textsc{MIS} if and only if $(H, k, I_s, I_t)$ is a yes-instance of \textsc{ISR-TS}; before any token can slide to $R$ we must have $k$ tokens in $G$ forming a multicolored independent set.

\subsubsection{Galactic graphs.}
While the parameterized complexity of $\textsc{ISR-TJ}$ on sparse classes of graphs is well-understood, the situation is quite different for $\textsc{ISR-TS}$. Indeed, $\textsc{ISR-TS}$ is more complicated than the token jumping variant even in sparse classes of graphs because of what is called the \emph{bottleneck effect}~\cite{DBLP:journals/corr/abs-2204-05549}. Under the token jumping model, the existence of a large independent set in the non-neighborhood of both the source and target independent sets is enough to ensure the existence of a reconfiguration sequence from $I_s$ to $I_t$. To the contrary, under the token sliding model, a small cut might prevent us from finding such a transformation.
Let us illustrate that behavior on the simple example of a star to which we attach a long path. If there are at least two tokens on the leaves of the star, none of the tokens on leaves can slide. Consequently, we will not be able to move any token from leaves of the star to the path.

In an ideal world, we would like to like to determine if there exist frozen tokens (in the sense that they will never able to slide) and remove them from the graph. While it can be done efficiently for trees~\cite{DBLP:conf/isaac/DemaineDFHIOOUY14}, this problem is hard in general. 
So one needs to find another strategy. Let us return to the example of the star to which a path is attached. One can easily notice that, if the path is long enough, we can form an equivalent instance by reducing its length. 
Note that by doing so we found 1) a large subset of vertices which can be replaced by a smaller one to form an equivalent instance and 2) the resulting graph was in the same class (in this case, trees).
In order to prove the existence of a fixed-parameter tractable algorithm, one usually wants to perform such reductions but, even if it is often easy to find a subset of vertices that behave ``nicely'' (in the sense that we understand quite well how the tokens behave in that subset), it is not always easy to reduce the size of the graph and obtain a resulting graph within the same class. Moreover, the proofs are usually technical since, quite often, neither of the directions of the equivalence are trivial to prove.
To overcome this issue, a more general version of $\textsc{ISR-TS}$ was introduced on galactic graphs~\cite{DBLP:journals/corr/abs-2204-05549}.

A \emph{galactic graph} is a graph where $V(G)$ is partitioned into two sets: the \emph{planets} 
and the \emph{black holes}.
In a galactic graph, the rules of the $\textsc{ISR-TS}$ game are slightly modified. When a token reaches a black hole, the token is
\emph{absorbed} by the black hole. 
Since black holes are considered to make tokens ``disappear'', it is allowed for tokens to be assigned to adjacent vertices as long as at least one of the vertices is a black hole. Furthermore, each black hole is allowed to absorb up to $k$ tokens. In addition, a black hole can also \emph{project} any of the tokens it previously absorbed onto any vertex in its neighborhood, be it a planet or a black hole.

Why galactic graphs? At first glance, they might seem very artificial. However, in practice, one often finds a large connected structure on which one can prove that one can ``hide'' as many tokens as one wants. However, it is usually complicated to prove the existence of a smaller structure with the same property while staying in the same class. Shrinking all this structure into a single vertex drastically simplifies this step as well as the technicalities of the proofs. 
An important result one can prove using black holes is the following:

\begin{lemma}
Let $G$ be a galactic graph and $I_s,I_t$ be two independent sets of size~$k$. If $G$ contains a shortest path $P$ of length $\Oof(k)$ such that $N_G(V(P)) \setminus V(P)$ does not contain any token of $I_s \cup I_t$, then $P$ can be contracted into a black hole.
\end{lemma}

Using this rule, together with other simple rules, one can prove the following:

\begin{theorem}\label{thm:ISR-TS_degree}
$\textsc{ISR-TS}$ is fixed-parameter tractable parameterized by $k+\Delta$, where $\Delta$ is the maximum degree of $G$.
\end{theorem}

\cref{thm:ISR-TS_degree} implies in particular that  $\textsc{ISR-TS}$ is fixed-parameter tractable on classes of graphs of bounded degree (and on classes of graphs of bounded bandwidth). To extend results to more general classes of graphs, we need more tools.

\subsubsection{Types.}
A surprisingly hard question is the following: Let $G$ be a graph, $I_s,I_t$ be two independent sets of size $k$ and $X$ be a subset of vertices such that $G - X$ contains many connected components. Is it possible to remove one of the components while preserving the existence of a transformation from $I_s$ to $I_t$? 

Since a token might perform an arbitrarily long walk in a component, it is not simple to prove that one of these components can be deleted to form an equivalent instance. Bartier et al.~\cite{DBLP:journals/corr/abs-2204-05549} introduced the notion of \emph{types} of walks in a component $H$ of $G - X$. The intuition behind the proof is that in a walk $W$ of a token $t$ in $H$, we can find a subset of important vertices $x_1,\ldots,x_r$, called {\em conflict vertices}, such that we can express $W$ as $x_1P_1x_2P_2\ldots,P_{r-1}x_r$, where $P_i$ is the path of $W$ linking $x_i$ to $x_{i+1}$. Then, in the walk $W$ performed by $t$, the only important information is $N(x_i) \cap X$ and $\cup_{u \in P_i} N( u) \cap X$. Thus, if we can prove that the number of conflict vertices in every walk $W$ is bounded with respect to $k$ and $|X|$, the (potentially unbounded length of) information contained in a walk $W$ can be summarized in $f(k,|X|)$ neighborhoods in $X$. The authors are then able to prove the following:

\begin{lemma}\label{lem:isrtj_multi}
Let $G$ be a graph, $X$ be a subset of vertices, and $I_s,I_t$ be two independent sets of $G$ of size $k$.
Let $\mathcal{S}$ be a transformation from $I_s$ to $I_t$ that minimizes the number of token slides involving a vertex of $X$. Almost all the components~$H$ of $G - X$ are well-behaved, i.e., the number of conflict vertices of a walk of a token $t$ in $H$ is bounded by a function of $k$ and $|X|$. 
\end{lemma}


The proof requires showing that if a walk $W$ of a token $t$ in $H$ has too many conflict vertices, then we should have, at some point, projected a token $t' \ne t$ on a component of $G - X$ where we can mimic the behavior of the walk $W$ of $t$ in order to decrease the number of token slides involving a vertex of $X$. 
Since, for~\cref{lem:isrtj_multi}, we can determine the well-behaved components of $G - X$ in \FPT-time and since the number of types of walks is bounded, if $G - X$ contains too many components, we can safely remove one of them.

\subsubsection{Applications and open questions.} Using these ingredients, i.e., galactic graphs and types, Bartier et al.~\cite{DBLP:journals/corr/abs-2204-05549} proved that $\textsc{ISR-TS}$ is fixed-parameter tractable on graphs of bounded degree, planar graphs, and chordal graphs of bounded clique number. The main ingredient of these proofs entails proving that, using the multi-component reduction described as well as additional reduction rules, we can reduce the maximum degree to $f(k)$. Then~\cref{thm:ISR-TS_degree} ensures that $\textsc{ISR-TS}$ is fixed-parameter tractable for all these classes.
Note that the idea of reducing the degree has also been used in~\cite{DBLP:journals/algorithmica/BartierBDLM21} to obtain fixed-parameter tractable algorithms for $\textsc{ISR-TS}$ on graphs with girth constraints. 
The authors in~\cite{girthfive} also used these ingredients to show that $\textsc{ISR-TS}$ is fixed-parameter tractable on graphs of girth five or more (the problem is \WO-hard on graphs of girth four or less~\cite{DBLP:journals/algorithmica/BartierBDLM21}). 

Natural next questions to consider are the following:

\begin{question}
Is $\textsc{ISR-TS}$ parameterized by $k$ fixed-parameter tractable on graphs of bounded treewidth? On minor-free graphs? On $d$-degenerate graphs? 
\end{question}

\section{Reconfiguration of (connected) dominating sets}\label{sec-dsr}

\subsection{\textsc{DSR-TJ} and \textsc{CDSR-TJ}}

\subsubsection{Reconfiguration of dominating sets.} On general graphs,
$\textsc{DSR-TJ}$ is \WT-hard parameterized by solution size $k$ plus the length of a reconfiguration sequence~$\ell$~\cite{DBLP:journals/algorithmica/MouawadN0SS17}. When parameterized by $\ell$ alone the problem is fixed-parameter tractable on any class where first-order model-checking is fixed-parameter tractable, the most general known cases being nowhere dense classes~\cite{DBLP:journals/jacm/GroheKS17} and classes of bounded twinwidth (assuming a contraction sequence is also given as part of the input)~\cite{DBLP:journals/jacm/BonnetKTW22}. To demonstrate why, we formulate the 
\textsc{DSR-TJ} problem as follows. We encode the instance 
$(G,k,D_s,D_t)$ as a colored graph, where the vertices of 
$D_s$ and $D_t$ are marked by two unary predicates, so that
they become accessible to first-order logic. First-order logic now
existentially quantifies the at most $\ell$ vertices that are changed in a reconfiguration sequence of length $\ell$. Now, it remains to verify that the vertices marked with the first predicate modified by the quantified changes form a dominating set. We note that fixed-parameter tractability using first-order model-checking for parameter $\ell$ also holds for $\textsc{ISR-TJ}$,  $\textsc{ISR-TS}$, and $\textsc{DSR-TS}$. 


Unfortunately, the approach via first-order model-checking does not immediately help us to deal with the parameter $k$. For the parameter $\ell$, we can find a fixed sentence expressing the existence of a reconfiguration sequence of length~$\ell$. However, when restricted to the parameter $k$ alone, because reconfiguration sequences can be arbitrary long, it is not possible to state the existence of such a sequence in first-order logic. 

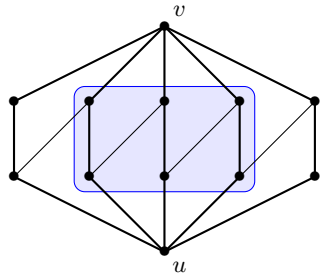
\begin{figure}[ht]
\begin{center}
\begin{tikzpicture}
\node (a1) at (0,0) {$\bullet$};
\node at (0.2,-0.2) {$u$};
\node (a2) at (0,3) {$\bullet$};
\node at (0.2,3.2) {$v$};
\draw [draw=blue,rounded corners,fill=blue!10!white] (-1.2, 0.8) rectangle (1.2,2.2);
\foreach \x in {-2cm,-1cm,0cm,1cm,2cm}
{
	\node at (\x,1) {$\bullet$};
	\node at (\x,2) {$\bullet$};	
	\draw[-,thick] (0,3) -- (\x,2) -- (\x,1) -- (0,0);
}
\draw[-] (-2,1) -- (-1,2);
\draw[-] (-1,1) -- (0,2);
\draw[-] (0,1) -- (1,2);
\draw[-] (1,1) -- (2,2);

\end{tikzpicture}
\end{center}
\caption{A graph with a dominating set of size $2$ and a $2$-domination core marked by a light-blue box with rounded corners. The only way to dominate the core by $2$ is to pick vertices $u$ and $v$.}
\label{fig:dom-core}
\end{figure}

Instead, the key tool to tackle \textsc{DSR-TJ} is based on the notion of domination cores. A \emph{$k$-domination core} in a graph $G$ is a subset $Y\subseteq V(G)$ such that every set of size at most $k$ dominating $Y$ also dominates the whole graph (see~\cref{fig:dom-core}).

For a graph $G$, we fix source and target dominating sets $D_s$ and $D_t$ of size~$k$ and a $k$-domination core $Y$. We let $R$ denote a set containing one vertex from each projection class of $\mathcal{C}_Y$. We call a vertex in $R$ a {\em representative} of its class. 

Let $H$ be the graph induced by the vertices of $Y\cup D_s\cup D_t\cup R$. Observe that 
$N_G(v)\cap Y=N_H(v)\cap Y$ for all $v\in V(H)$. For $v\in V(G)$, let $v_H=v$ if $v\in D_s\cup D_t\cup Y$ and otherwise let $v_H$ be the vertex of $R$ representing the projection class of~$v$. For $w\in V(H)\setminus (Y\cup D_s\cup D_t)$ fix an arbitrary vertex $w_G\in V(G)$ with 
$N_H(w)\cap Y=N_G(w_G)\cap Y$. 


\begin{lemma}
If $D_G\subseteq V(G)$ is a dominating set of $G$, then $D_H=\{v_H~:~v\in D_G\}$ is a dominating set of $H$. Conversely, if $D_H\subseteq V(H)$ is a dominating set of size at most $k$ of $H$, then $D_G=\{w_G~:~w\in D\}$ is a dominating set of $G$. 
\end{lemma}


The first statement of the lemma follows from $H$ being an induced subgraph of~$G$.
For the second statement, since $D_H$ dominates $H$, it in particular dominates~$Y$ in $H$. Because $H$ is an induced subgraph of $G$, the set $D_G$ dominates~$Y$ also in $G$. By definition of a $k$-domination core, $D_G$ dominates $G$. 

\smallskip
The use of $k$-domination cores for \textsc{DSR-TJ} is now  immediate. 

\begin{corollary}\label{lem:core-sequence}
There exists a reconfiguration sequence 
from $D_s$ to $D_t$ in~$G$ 
if and only if there exists a reconfiguration sequence from $D_s$ to $D_t$ in~$H$. 
\end{corollary}

Since there are at most $2^{|C|}$ different projection classes,
it follows that the graph $H$ is small in relation to $|C|$. Even better bounds can be achieved when the VC-dimension of the graph class under consideration is bounded~\cite{DBLP:journals/jct/Sauer72,pjm/1102968432}. 

\begin{lemma}
The graph $H$ has at most $2k+|C|+2^{|C|}$ vertices. Furthermore, when the VC-dimension of $G$ is bounded by $d$, then $H$ has at most
$\Oof(k+|C|^d)$ vertices. 
\end{lemma}

Now it is easy to derive fixed-parameter tractable algorithms for all classes of graphs that admit small domination cores. 

\begin{theorem}
Assume $\Cc$ is a class of graphs such that for every $G\in \Cc$ and for every $k\in \mathbb{N}$ there exists a polynomial-time computable $k$-domination core of size $f(k)$, for some computable function $f$. Then \textsc{DSR-TJ} is fixed-parameter tractable on $\Cc$. Furthermore, when $f$ is a polynomial function and $\Cc$ has bounded VC-dimension, then $\textsc{DSR-TJ}$ admits a polynomial kernel on $\Cc$.
\end{theorem}

Domination cores were first introduced by Dawar and Kreutzer~\cite{DBLP:conf/fsttcs/DawarK09} in their study of the \textsc{Distance-$r$ Dominating Set} problem on nowhere dense classes. When~$k$ is the size of a minimum dominating set, then linear $k$-domination cores (i.e.\ cores of size $\Oof(k)$) are known to exist on classes with bounded expansion~\cite{DBLP:conf/stacs/DrangeDFKLPPRVS16} and polynomial (and in fact almost linear) cores exist on nowhere dense classes \cite{DBLP:journals/talg/KreutzerRS19,DBLP:conf/icalp/EickmeyerGKKPRS17}. When $k$ is not minimum, in all of these cases $k$-domination cores are known to be of polynomial size. The most general classes of graphs known to admit domination cores are semi-ladder-free classes~\cite{DBLP:conf/stacs/FabianskiPST19}. These classes include biclique-free classes, on which the cores are of polynomial size (with the exponent depending on the size of the excluded biclique) and which have bounded VC-dimension. Their use for reconfiguration was first observed in~\cite{DBLP:journals/jcss/LokshtanovMPRS18}. 

\begin{corollary}
\textsc{DSR-TJ} is fixed-parameter tractable on every semi-ladder-free class of graphs. \textsc{DSR-TJ} admits a polynomial kernel on every biclique-free class of graphs. 
\end{corollary}

\begin{question}
Does $\textsc{DSR-TJ}$ admit a linear kernel on planar graphs?
\end{question}


The methods based on domination cores are in fact limited to classes with bounded co-matching index (which in particular have bounded semi-ladder index). It is easily seen that the graph consisting of two cliques $A,B$ of size $n$ such that the edges between $A$ and $B$ induce a co-matching has a dominating set of size $2$ but does not contain a non-trivial $2$-domination core. For example, while it is known that the \textsc{Dominating Set} problem parameterized by $k$ is fixed-parameter tractable on every class of bounded cliquewidth it is not known whether this is true for \textsc{DSR-TJ}. 

\begin{question}
Is $\textsc{DSR-TJ}$ parameterized by $k$ fixed-parameter tractable on graphs of bounded cliquewidth? 
\end{question}


The results of this section naturally generalize to the distance-$r$ version of \textsc{Dominating Set}. A distance-$r$ dominating set in a graph $G$ is a dominating set in the $r$th power $G^r$ of $G$. The $r$th power $G^r$ of an undirected graph $G$ is another graph that has the same set of vertices, but in which two vertices are adjacent when their distance in $G$ is at most $r$. Observe that for $r\geq 2$ the $r$th powers of classes with unbounded degree contain arbitrarily large cliques. However, for example powers of nowhere dense graphs still have bounded semi-ladder index~\cite{DBLP:conf/stacs/FabianskiPST19}, and hence, \textsc{DSR-TJ} is fixed-parameter tractable on powers of nowhere dense classes. One can obtain almost linear kernels when the underlying nowhere dense graph is given, or equivalently, when considering the reconfiguration variant of the  \textsc{Distance-$r$ Dominating Set} problem on nowhere dense classes~\cite{DBLP:journals/combinatorics/Siebertz18}. 

\subsubsection{Reconfiguration of connected dominating sets.}
\begin{figure}[ht]
\begin{center}
\begin{tikzpicture}
\node (a1) at (0,0) {$\bullet$};
\node (a2) at (0,3) {$\bullet$};
\foreach \x in {-2cm,-1cm,0cm,1cm,2cm}
{
	\node at (\x,1) {$\bullet$};
	\node at (\x,2) {$\bullet$};	
	\draw[-,thick] (0,3) -- (\x,2) -- (\x,1) -- (0,0);
}
\draw[-] (-2,1) -- (-1,2);
\draw[-] (-1,1) -- (0,2);
\draw[-] (0,1) -- (1,2);
\draw[-] (1,1) -- (2,2);

\filldraw[fill=black!50] (0,0) circle (3pt);
\filldraw[fill=black!50] (0,3) circle (3pt);
\filldraw[fill=black!50] (-2,1) circle (3pt);
\filldraw[fill=black!50] (-2,2) circle (3pt);

\filldraw[fill=white!90] (0.2,0) circle (3pt);
\filldraw[fill=white!90] (0.2,3) circle (3pt);
\filldraw[fill=white!90] (2,1) circle (3pt);
\filldraw[fill=white!90] (2,2) circle (3pt);


\node at (3,1.5) {$\Rightarrow$};

\begin{scope}[xshift=6cm]
\node (a1) at (0,0) {$\bullet$};
\node (a2) at (0,3) {$\bullet$};
\foreach \x in {-2cm,-1cm,1cm,2cm}
{
	\node at (\x,1) {$\bullet$};
	\node at (\x,2) {$\bullet$};	
	\draw[-,thick] (0,3) -- (\x,2) -- (\x,1) -- (0,0);
}
\draw[-] (-2,1) -- (-1,2);
\draw[-] (-1,1) -- (1,2);
\draw[-] (1,1) -- (2,2);
\end{scope}
\end{tikzpicture}
\end{center}
\caption{On the left: a graph showing that no induced subgraph kernels exist. On the right: how to reduce the instance.}
\label{fig:final-reduction}
\end{figure}
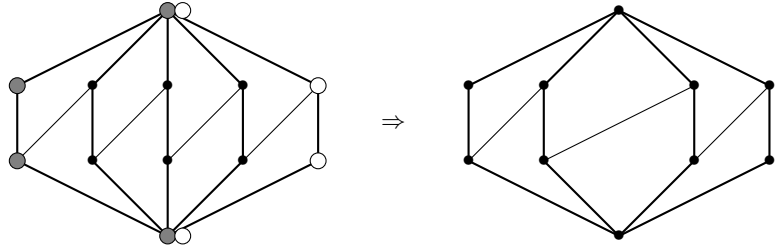

The connected version of the problem, \textsc{CDSR-TJ}, is much harder. Lokshtanov et al.~\cite{DBLP:journals/algorithmica/LokshtanovMPS22} showed that the problem is hard parameterized by $k+\ell$ even on the class of $5$-degenerate graphs. Even though there exist small domination cores on this class, a key difference is that with the additional connectivity constraint it is no longer sufficient to keep one representative vertex of each projection class in addition to the core (and source and target solutions). This difficulty is highlighted in the following example, showing that in fact one cannot expect to find kernels that consist of induced subgraphs even on planar graphs. \cref{fig:final-reduction} depicts a graph with connected dominating sets $D_s$ (in gray) and $D_t$ (in white). The connectivity constraint forces that there is only one possibility for reconfiguration, consisting of the step-by-step reconfiguration of the middle two vertices from left to right. If we delete any vertex not in $D_s\cup D_t$, then a reconfiguration sequence can no longer exist.  


Further, the authors~\cite{DBLP:journals/algorithmica/LokshtanovMPS22} showed that the problem is fixed-parameter tractable and in fact admits a polynomial kernel on planar graphs. This result again builds on domination cores. Given a planar graph, as for \textsc{DSR-TJ}, we compute a small $k$-domination core. We then use planarity to identify irrelevant vertices (vertices that can be removed without changing the reconfiguration properties of the graph) and remove them. This finally leaves us with the only difficulty, which is shown in~\cref{fig:final-reduction}. By analyzing the interactions with the domination core and using planarity, we identify large subgraphs of the depicted form, which can then be replaced by constant-size gadgets such that the reconfiguration properties of~$G$ are preserved.  Overall, this leads to a polynomial kernel. 

\begin{question}
Does $\textsc{CDSR-TJ}$ admit a linear kernel on planar graphs?
\end{question}

Unfortunately, the methods discussed above are not generic but are rather hand-crafted for planar graphs. The following remains a fundamental open question:

\begin{question}
Is $\textsc{CDSR-TJ}$ parameterized by $k$ fixed-parameter tractable on graphs of bounded pathwidth? 
\end{question}

\subsection{\textsc{DSR-TS} and \textsc{CDSR-TS}}
To the best of our knowledge, the parameterized complexity of \textsc{DSR-TS} and \textsc{CDSR-TS} has not been studied so far. However, one can immediately see that for the simple case of graphs of bounded degree we obtain fixed-parameter tractable algorithms, since a graph $G$ of maximum degree $\Delta$ having a (connected) dominating set of size at most $k$ can have at most $k\Delta$ vertices. However, for all the other classes considered in this survey, the parameterized complexity of both problems is still open. 


\begin{question}
Is $\textsc{DSR-TS}$ parameterized by $k$ fixed-parameter tractable on graphs of bounded pathwidth/treewidth? On planar graphs?
\end{question}

\begin{question}
Is $\textsc{CDSR-TS}$ parameterized by $k$ fixed-parameter tractable on graphs of bounded pathwidth/treewidth? On planar graphs?
\end{question}

\section{Conclusion and further open problems}\label{sec-conclusion}
In some aspects, the study of the parameterized complexity of reconfiguration variants of \textsc{Independent Set} and \textsc{(Connected) Dominating Set} parameterized by the number of tokens is still ``catching up'' with what is known for the underlying problems when parameterized by solution size. While some techniques are new and focused on reconfiguration, other tools are adaptions of tools that were originally developed for solving the underlying problems. It is our hope that this line of research will eventually lead to results migrating in the opposite direction. That is, insights from the reconfiguration variant of a problem will lead to new insights for solving the underlying problem (or at least better understanding its complexity). Getting there will of course require solving the remaining open questions and hopefully generalizing these results to meta-theorems. We note that our list of open questions is in no way exhaustive. Our questions are focused on understanding the boundary between tractability and intractability. This boundary is still not very clear for almost all of the reconfiguration problems we considered. 

Finally, we note that many different interesting parameterizations of reconfiguration problems are possible and almost no work has been done in this direction (except for a few results for parameter $\ell$). Moreover, the reconfiguration rules considered so far, i.e., jumping and sliding, are in no way unique. One can also consider many different general rules like for instance the movement of several tokens simultaneously.  

\bibliographystyle{plain}
\bibliography{references}

\end{document}